# Further Evidences for Spin Glass like behavior in NiO nanoparticles

S. D. Tiwari and K. P. Rajeev

Department of Physics, Indian Institute of Technology, Kanpur 208016, Uttar Pradesh,

India

Nickel oxide nanoparticles are prepared by a sol gel method and characterized by x-ray diffraction and transmission electron microscope. Here we present measurements on temperature and field dependence of magnetization and time dependence of thermoremanent magnetization. Our conclusion based on these measurements is that the system shows spin glass like behavior.

## I. Introduction

Work on nanoparticles has become of increasing interest because of new and interesting properties that the material may show when the grain size is reduced to a few nanometer. If the surface to volume ratio, which varies as the reciprocal of particle size, for antiferromagnetic particles becomes sufficiently large then the particles can have nonzero net magnetic moment because of uncompensated spins at the surface. In one of our recent work[1] mainly based on ac susceptibility measurements, we claimed that the behavior of NiO nanoparticles is spin glass like. Here we report experiments on same sample used in the earlier work.

## II. Experimental results and discussions

Figure 1 shows that the peak temperature in susceptibility vs. temperature curve decreases as particle size increases. It shows clearly that the behavior of the system is not superparamagnetic because superparamagnetic blocking temperature increases with increase in particle size.

We measured the ac susceptibility at various bias fields H with an ac field of 1 G amplitude and 10 Hz frequency. This is shown in figure2. As H increases, the peak temperature decreases and the peak broadens. We find that the peak temperature $T_f$ decreases as H increases following a power law dependence, namely, $\delta T_f \propto H^{2/3}$ (see inset of figure2). This dependence corresponds to the so called the de Almeida-Thouless (AT) line[2] given by: $H \propto (1 - T/T_f)^{3/2}$. The extrapolation of the AT line back to H = 0 gives the spin glass transition temperature $T_f$.

Thermoremanent magnetization (TRM) is the time dependent remanent magnetization obtained when a sample is cooled from $T>T_f$ to a temperature $T<T_f$ in a field H and subsequently the field is removed. In this work for measuring TRM the sample is cooled from 300K to the temperature of interest in a 100 G magnetic field.

Expressions like stretched exponential $M(t) \sim \exp(-(t/\tau)^\beta)$, power law $M(t) \sim t^{-\beta}$, logarithmic dependence $M(t) \sim \ln(t/\tau)$ etc. have been popularly used for describing magnetization relaxation of magnetic materials. These are just empirical relations. Unfortunately these are also mathematical approximations of a number of physical models. Therefore agreement with these expressions tells us nothing about the physical mechanism of response of the material. Fig3 shows TRM at a few temperatures. The decay is found to be well described by the relation $M(t) = M_0 + S \ln(t)$. This relation have been used to describe the time dependence of thermoremanent magnetization of different spin glasses[3].

## III. Conclusion

The behavior of the sol gel prepared NiO nanoparticles is spin glass like rather than being superparamagnetic.

Figures

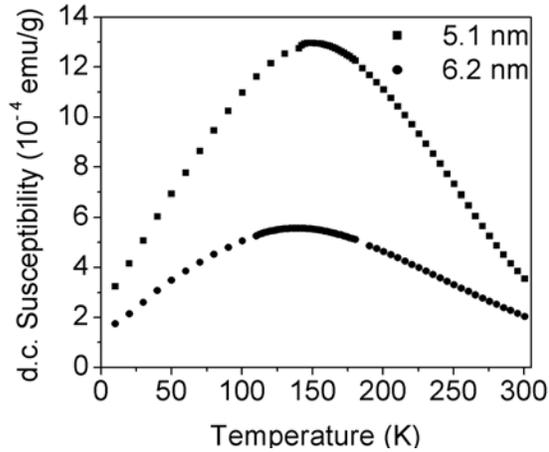

Fig. 1. Low field (100 G) zero field cooled susceptibility of NiO nanoparticles of two different sizes. Figure shows a decrease in peak temperature as the particle size increases. Obviously this is not a case of superparamagnetism.

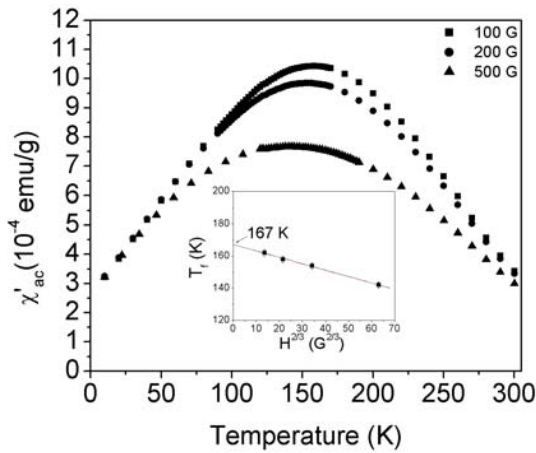

Fig.2. Field dependence of ac susceptibility vs. temperature curve. Inset shows field dependence of the spin glass transition temperature $T_f$ showing the AT line, i.e. $\delta T_f \propto H^{2/3}$. $T_f$ (H=0) is obtained by extrapolating the AT line back to H = 0.

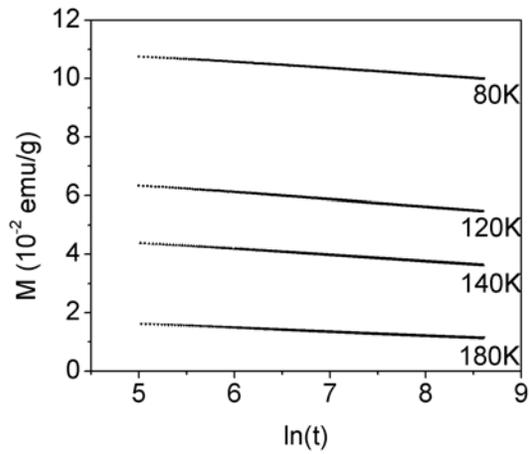

Fig.3. Time dependence of Thermoremanent magnetization. The magnetization changes linearly with ln (t).

---

[1] www.arXiv.org/cond-mat/0408427.

[2] B. Martinez, et al., Phys. Rev. Lett. **80**, 181 (1998); A. Fukaya, et al., J. Appl. Phys. **89**, 7053 (2001); S. Dhar, et al., Phys. Rev. B **67**, 165205 (2003).

[3] C. N. Guy, J. Phys. F: Metal Phys. **8**, 1309 (1978); S. H. Chun, et al., J. Appl. Phys. **90**, 6307 (2001); S. Dhar, et al., Phys. Rev. B **67**, 165205 (2003).